\documentstyle[preprint,aps]{revtex}
\tighten
\begin{document}
\draft
\title{Critical test of multi-{\it j} supersymmetries from
magnetic moment measurements}
\author{Serdar Kuyucak}
\address{Department of Theoretical Physics,
Research School of Physical Sciences and Engineering,\\
Australian National University, Canberra, ACT 0200, Australia}
\author{Andrew E. Stuchbery}
\address{Department of Nuclear Physics,
Research School of Physical Sciences and Engineering,\\
Australian National University, Canberra, ACT 0200, Australia}
\date{\today}
\maketitle
\begin{abstract}
Magnetic moment measurements in odd nuclei directly probe the distribution
of fermion states and hence provide one of the most critical tests for
multi-$j$ supersymmetries in collective nuclei.
Due to complexity of calculations and lack of data, such tests have
not been performed in the past.
Using the Mathematica software, we derive analytic expressions for
magnetic moments in the $SO^{(BF)}(6) \times SU^{(F)}(2)$ limit of
the $U(6/12)$ supersymmetry and compare the results with recent
measurements in $^{195}$Pt.
\end{abstract}

The first examples of supersymmetry were found in the spectra of collective
nuclei in the framework of the interacting boson-fermion model (IBFM)
\cite{iac91}, and have attracted considerable attention during the past decade
(see Refs. \cite{iac91,ver87} for reviews).
The single-$j$ supersymmetries, which were first proposed \cite{iac80,bal81},
coupled a fermion in a single specific orbit to a collective boson core.
As there are usually several single-particle orbitals available for the
fermion,
these models had limited success \cite{ver87}.
This restriction was overcome in multi-$j$ supersymmetries
\cite{bal83} through the introduction of the pseudo-spin mechanism which
allowed coupling of any number of single-particle orbits to the boson core.
In a given supersymmetry scheme, the wave functions are fixed from the outset
(i.e. independent of the Hamiltonian parameters) and require very specific
couplings of the fermion states to the boson states.
Since the wave functions are independent of the Hamiltonian, energies are not
a sensitive test of the model, and one has to consider electromagnetic
($E2$ and $M1$) properties, and one- and two-nucleon transfer reactions.
Of these, the $E2$ transition rates are not very sensitive to the single
particle distributions because i) they are dominated by the boson contribution,
the fermion contribution to the $E2$ matrix element (m.e.) being $1/N$
smaller than the boson part where $N$ is the boson number, and
ii) the fermion effective charges are all taken to be equal
and do not distinguish between different orbitals.
One-nucleon transfer reactions are sensitive to the single
particle distributions, but the transfer operator contains up to two
free parameters for each $j$-orbital which allows too much flexibility
to provide a definitive test.

In contrast, the $M1$ properties are free of these shortcomings, namely,
i) the boson and fermion contributions to the $M1$ m.e. are similar
(in fact, the latter are usually larger), ii) the g-factors of all
orbitals differ significantly, and iii) the $M1$ operator for odd
nuclei does not contain any free parameters.
Thus magnetic moments and $B(M1)$ values offer one of the most critical
tests for probing the coupling schemes predicted by multi-$j$ supersymmetries.
Such tests, with one limited exception \cite{bij88}, have not been performed
in the past due to the complexity of calculations and the lack of data.
In this Letter, we point out that the algebraic computations
that were previously deemed too complicated can be performed relatively easily
using the Mathematica software \cite{wol91}.
We derive analytic expressions for magnetic moments in the
$SO^{(BF)}(6) \times SU^{(F)}(2)$ limit of the $U(6/12)$ supersymmetry
and compare the results with a new extensive set of data \cite{lam93} in
$^{195}$Pt which, together with $^{194}$Pt, furnishes one of the best
known examples of this supersymmetry.

The wave functions in the $SO^{(BF)}(6) \times SU^{(F)}(2)$ dynamical
symmetry can be expanded in terms of the boson-fermion product states as

\begin{eqnarray}
&&|[N],[N_1,N_2],(\sigma_1,\sigma_2,\sigma_3),(\tau_1,\tau_2),\gamma,L,J
\rangle \nonumber\\
&&= \sum_{\sigma_B,\tau_B,\gamma_B,L_B,j}
\alpha(\sigma_B,\tau_B,\gamma_B,L_B,j;J)
\left[|\sigma_B,\tau_B,\gamma_B,L_B\rangle \times |j\rangle\right]^{(J)}.
\end{eqnarray}
Here the quantum numbers $N,\sigma,\tau$ label the $U(6), O(6), O(5)$
groups respectively, and $j$ denotes the $p_{1/2}, p_{3/2}, f_{5/2}$
single-particle orbits.
The expansion coefficients $\alpha$ are given in terms of the
isoscalar factors $\xi$ for the group chain
$U(6) \supset O(6) \supset O(5) \supset O(3)$ as \cite{van84,bij85}

\begin{equation}
\alpha(\sigma_B,\tau_B,\gamma_B,L_B,j;J) =
(-)^{L_B+J+1/2}\hat{L}\hat{j}\left\{ \begin{array}{ccc} L_B & L_F & L\\
1/2 & J & j\end{array}\right\}
\xi^{[N],(\sigma_B,0,0),(\tau_B,0),L_B}
_{[N_1,N_2],(\sigma_1,\sigma_2,\sigma_3),(\tau_1,\tau_2),L},
\end{equation}
where the curly bracket denotes a $6-j$ symbol and $\hat{j}=\sqrt{2j+1}$.

Introducing the boson and fermion creation (annihilation) operators
$d^\dagger_\mu$, $a^\dagger_{j\mu}$ ($d_\mu$, $a_{j\mu}$),
the $M1$ operator in the IBFM is given by

\begin{equation}
T(M1) = q_1 [d^\dagger \tilde d]^{(1)} + \sum_{jj'}t_{1jj'}
[a_j^\dagger {\tilde a_{j'}}]^{(1)},
\end{equation}
where tilde denotes ${\tilde a}_{jm}=(-)^{j-m} a_{j,-m}$.
The parameters $q_1$ and $t_{1jj'}$ in Eq. (3) are determined from
the boson g-factor and single-particle matrix elements respectively.
In the calculation of magnetic moments, in general, the cross terms
with $j=l\pm1/2$ can also contribute to the matrix elements.
However, in the case of the $SO^{(BF)}(6) \times SU^{(F)}(2)$ dynamical
symmetry, the $p_{1/2}$ and $p_{3/2}$ single particle states couple to
different boson states, and hence such cross terms all vanish.
This effectively simplifies the magnetic moment operator to diagonal terms
only  which we will rewrite in terms of the boson and single-particle angular
momentum operators as

\begin{eqnarray}
&&{\bf\hat\mu} = g_B {\bf L}_B + \sum_j g_j {\bf L}_j, \nonumber\\
&&{\bf L}_B = \sqrt{10} [d^\dagger {\tilde d}]^{(1)},  \nonumber\\
&&{\bf L}_j = -\left[ j(j+1)(2j+1)/3\right]^{1/2}
[a_j^\dagger {\tilde a}_j]^{(1)}.
\end{eqnarray}
The g-factor for bosons, $g_B$, is determined from the supersymmetric even-even
partner, and for (neutron) fermions from the single-particle (Schmidt) values
as
\begin{equation}
g_j=\pm g_s/(2l+1)~\text{for}~j=l\pm 1/2,
\end{equation}
where $g_s$ is the spin g-factor of neutrons which, with the standard
quenching factor of 0.6, has the value $g_s=0.6g_s^{free}=-2.3$.
The expectation value of the operator (4) in the states (1)
can be calculated easily and gives for the magnetic moments

\begin{equation}
\mu_J = \sum_{\sigma_B,\tau_B,\gamma_B,L_B,j}
\left(\alpha(\sigma_B,\tau_B,\gamma_B,L_B,j;J)\right)^2
{1\over2(J+1)}\left[ (\bar J+\bar L_B-\bar j)g_B +
(\bar J-\bar L_B+\bar j)g_j\right]
\end{equation}
Here the bar denotes $\bar J=J(J+1)$. The advantages of magnetic moment
measurements for testing the multi-$j$ supersymmetries noted above, are evident
from Eq. (6). For example, the magnetic moments are sensitive to the occupation
of the fermion states because the $g_j$ values differ significantly. If all the
g-factors were identical (c.f. the usual assumption
for E2 effective charges), i.e.
$g_j=g_B=g$, then one would obtain $\mu_J=gJ$, which gives no information about
the single-particle distribution.

Using the isoscalar factors given in Refs. \cite{van84,bij85},
magnetic moments can be evaluated in a straightforward manner from Eq. (6).
However, the algebraic manipulations required are very tedious and
lengthy, and they have not been calculated in the past.
We have overcome this problem by employing the Mathematica software
\cite{wol91} in the evaluation of Eq. (6)
and obtained relatively simple expressions for the magnetic moments.
As an example, we show an intermediate and the final step in the
calculations for the ground-band states,
$|[N],[N+1,0],(N+1,0,0),(\tau,0),2\tau,2\tau+1/2\rangle$

\begin{eqnarray}
\mu_{2\tau +1/2} = &&[20(N+1)(N+2)(2\tau+5)]^{-1}\bigl\{
40N(N+2)\tau(2\tau+5) g_B \nonumber\\
&&+ 5(N+\tau+4)(N-\tau+1)(2\tau+5) g_{1/2} \nonumber\\
&&- 2(N-\tau)(N-\tau+1)(6\tau+5) g_{3/2} \nonumber\\
&&+ \left( N^2(22\tau+35)+N(56\tau^2+202\tau+35)
+22\tau^3+213\tau^2+465\tau\right) g_{5/2}\bigr\}.
\end{eqnarray}
Results at this step were checked for errors by ensuring that when one uses
the same value $g$ for all g-factors, $\mu_J=gJ$.
Substituting $g_j$ from Eq. (5) gives the final result

\begin{equation}
\mu_{2\tau +1/2} = {2N\tau \over N+1} g_B
- {N^2(22\tau+35)+15N(6\tau+7)+8\tau^3+24\tau^2+108\tau+70
\over 42(N+1)(N+2)(2\tau+5)} g_s,
\end{equation}
which is not much more complicated than the corresponding expression for the
quadrupole moment. The results of calculations for all low-lying levels
of interest are listed in Tables I-III.

In the remainder, we compare the supersymmetry predictions
with recent g-factor measurements in $^{195}$Pt \cite{lam93},
one of the most studied and best examples of the $U(6/12)$ supersymmetry
\cite{bal83,van84,bij85,war82,sun83,ver83,bru85,mau86}.
Table IV compares the experimental g-factors with the theoretical ones
obtained from Tables I-III using $N=6$, $g_B=0.3$ (determined from
$^{194}$Pt \cite{stu91}), and $g_s=-2.3$.
In testing the quality of supersymmetry, the quantitative estimate
\[\phi=\left[ \Sigma |exp-th|/\Sigma |exp|\right] \%, \]
has often been used when discussing energy levels and transition rates
\cite{ver87}.
However, this is not very useful for the present comparisons of g-factors
because it takes no account of either experimental uncertainties or the level
of agreement that might reasonably be expected from a parameter-free
calculation. We prefer, therefore, a graphical presentation followed by a case
by case discussion.

We compare in Fig. 1 the experimental and theoretical g-factors for the
stretched states ($(\tau,0),L=2\tau$) in the $(N+1,0,0)$ and $(N,1,0)$
representations, for which there are complete sets of data up to spin
9/2. Aside from the ground-state, to be  discussed below, the agreement
is good, especially for the $(N,1,0)$ representation.
We emphasize that, in contrast to other tests, no free parameters are
used in the present g-factor calculations. In view of this, the level of
agreement is remarkable.

A significant deviation occurs for the ground state which is measured very
accurately.
Denoting the basis states by $|L_B\times j\rangle$, its decomposition
is given by
\begin{equation}
|1/2_1\rangle = \sqrt{5/8}~|0_1\times 1/2\rangle +
\sqrt{3/20}~|2_1\times 3/2\rangle - \sqrt{9/40}~|2_1\times 5/2\rangle.
\end{equation}
The g-factors for the three basis states in Eq. (9) are, in order,
0.77, 1.37 and 0.37.
The basis state $|2_1\times 3/2\rangle$ which has the largest g-factor,
has the smallest amplitude. Thus a possible explanation is that the ground
state has a larger $p_{3/2}$ component than predicted by the supersymmetry.
An alternative explanation may be the inadequacy of the simple
$M1$ operator used for bosons.
Such an operator cannot generate $M1$ transitions and therefore
has been modified with the addition of the two-body operator,
$[Q\times L]^{(1)}$.
The quadrupole operator, $Q$, has a vanishing m.e. in the ground
state, so this term could lead to very different contributions for the ground-
and other states. The calculation of m.e. for two-body operators is,
however, rather involved and this possibility needs to be pursued numerically.

For the remaining (non-stretched) states the data are too scant for
graphical presentation. Nevertheless, we discuss the two
$J=5/2$ states at 455 and 544 keV, both of which have
relatively large measured g-factors with large uncertainties. There are 13
possible basis states for $J=5/2$ with $L_B=2$ (twice), 3, 4(twice),
and $j=1/2, 3/2, 5/2$.
The average g-factor for all these unmixed states is 0.30, the highest
being 0.76 for the $|4\times 3/2\rangle$ state. All the other basis
states have g-factors around the average value or lower.
It is extremely unlikely that both of these 5/2 states are dominated
by the $|4\times 3/2\rangle$ configuration to the exclusion of many others.
In other words, $g_{5/2}\approx 0.6$ for these states would be
very difficult to explain in any particle-core model.
The supersymmetry model is, nevertheless, consistent with the experimental
results as the g-factors of these states may actually have
values near the lower limits allowed by the experimental uncertainties.

In conclusion, we have presented a critical test of the $U(6/12)$
supersymmetry from g-factor measurements in $^{194-195}$Pt.
The supersymmetry scheme makes definite predictions for the boson-fermion
wave functions which have previously been tested through level energies,
E2 transition rates and transfer reactions.
In comparison with these observables, g-factors provide
a more sensitive test of the wavefunctions because they depend directly
on the single-particle distributions.
Given that our g-factor calculations are parameter free, the level of
agreement between theory and experiment is remarkable and, in general,
supports the multi-$j$ supersymmetry model in $^{195}$Pt.
Further g-factor measurements and similar tests are planned for
other nuclei which evidence supersymmetry.

This work is supported in part by the Australian Research Council.

\begin{figure}
\caption{Comparison of g-factors for the stretched states
($(\tau_1,\tau_2)=(\tau,0),L=2\tau$) in the
$(N+1,0,0)$ and $(N,1,0)$ representations.}
\label{figure1}
\end{figure}

\mediumtext
\begin{table}
\caption{Magnetic moments for the states
$|[N],[N+1,0],(N+1,0,0),(\tau,0),L,J\rangle$.
The factor $f$ denotes $f=N(N+2)$.
\label{table1}}
\begin{tabular}{cccc}
$(\tau_1,\tau_2)$&$L$&$J$&$(N+1)(N+2)\mu_J$\\ \tableline
(0,0)&0&1/2&$-{1\over6}(N+1)(N+2)g_s$\\
(1,0)&2&3/2&${9\over5}fg_B+{1\over70}(11f+23N+70)g_s$\\
(1,0)&2&5/2&$2fg_B-{1\over98}(19f+27N+70)g_s$\\
(2,0)&2&3/2&${9\over5}fg_B+{1\over210}(17f+5N-38)g_s$\\
(2,0)&2&5/2&$2fg_B-{1\over882}(139f+115N+134)g_s$\\
(2,0)&4&7/2&${35\over9}fg_B+{1\over3402}(661f+1321N+5282)g_s$\\
(2,0)&4&9/2&$4fg_B-{1\over378}(79f+127N+446)g_s$\\
\end{tabular}
\end{table}

\mediumtext
\begin{table}
\caption{Magnetic moments for the states
$|[N],[N,1],(N,1,0),(\tau,0),L,J\rangle$.
The factor $f$ denotes $f=N(N+4)$.
\label{table2}}
\begin{tabular}{cccc}
$(\tau_1,\tau_2)$&$L$&$J$&$(N+1)(N+3)\mu_J$\\ \tableline
(1,0)&2&3/2&${9\over20}(f+11)g_B+{3\over35}(5f+3)g_s$\\
(1,0)&2&5/2&${1\over2}(f+11)g_B-{1\over294}(95f+141)g_s$\\
(2,0)&2&3/2&${9\over50}(7f+41)g_B+{1\over210}(7f+141)g_s$\\
(2,0)&2&5/2&${1\over5}(7f+41)g_B-{1\over882}(119f+597)g_s$\\
(2,0)&4&7/2&${7\over18}(7f+41)g_B+{1\over3402}(1211f-2967)g_s$\\
(2,0)&4&9/2&${2\over5}(7f+41)g_B-{1\over378}(119f-123)g_s$\\
\end{tabular}
\end{table}

\narrowtext
\begin{table}
\caption{Same as Table II but for the states with $(\tau_1,\tau_2)=(1,1)$.
\label{table3}}
\begin{tabular}{ccc}
$L$&$J$&$\mu_J$\\ \tableline
1&1/2&${1\over3}g_B-{1\over6}g_s$\\
1&3/2&${1\over2}g_B-{1\over10}g_s$\\
3&5/2&${10\over7}g_B+{59\over254}g_s$\\
3&7/2&${3\over2}g_B-{3\over14}g_s$\\
\end{tabular}
\end{table}

\begin{table}
\caption{Comparison of supersymmetry predictions from Tables I-III with the
experimental g-factors in $^{195}$Pt \protect \cite{lam93}.}
\label{table4}
\begin{tabular}{cccddd}
 & & & &\multicolumn{2}{c}{$g_J=\mu_J/J$}\\
$(\tau_1,\tau_2)$&$L$&$J$&$E_x(keV)$&experiment&theory\\
\tableline
\multicolumn{3}{l}{$(\sigma_1,\sigma_2,\sigma_3)=(7,0,0)$}& & & \\
(0,0)&0&1/2&0&1.22\tablenotemark[1]&0.77\\
(1,0)&2&3/2&211&0.10(2)\tablenotemark[1]&0.02\\
(1,0)&2&5/2&239&0.25(4)&0.40\\
(2,0)&2&3/2&525& &0.20\\
(2,0)&2&5/2&544&0.60(20)&0.35\\
(2,0)&4&7/2&613&0.41(12)&0.13\\
(2,0)&4&9/2&667&0.34(4)&0.35\\
\multicolumn{3}{l}{$(\sigma_1,\sigma_2,\sigma_3)=(6,1,0)$}& & & \\
(1,0)&2&3/2&99&-0.41(4)\tablenotemark[1]&-0.53\\
(1,0)&2&5/2&130&0.36(3)\tablenotemark[1]&0.36\\
(2,0)&2&3/2&420& &0.20\\
(2,0)&2&5/2&455&0.63(22)&0.30\\
(2,0)&4&7/2&508&0.16(2)&0.03\\
(2,0)&4&9/2&563&0.34(3)&0.35\\
(1,1)&1&1/2&222& &0.97\\
(1,1)&1&3/2&199& &0.25\\
(1,1)&3&5/2&389&0.16(4)&-0.04\\
(1,1)&3&7/2&450& &0.27\\
\end{tabular}
\tablenotetext[1]{From Ref. \cite{rag89}.}
\end{table}


\begin{references}
\bibitem{iac91} F. Iachello and P. Van Isacker, {\it The interacting
boson-fermion model} (Cambridge University Press, Cambridge, 1991).
\bibitem{ver87} J. Vervier, Riv. Nuovo Cimento {\bf 10}, No. 9 (1987).
\bibitem{iac80} F. Iachello, Phys. Rev. Lett. {\bf 44},772 (1980).
\bibitem{bal81} A.B. Balantekin, I. Bars, and F. Iachello,
Phys. Rev. Lett. {\bf 47},19 (1981); Nucl. Phys. {\bf A370},384 (1981).
\bibitem{bal83} A.B. Balantekin, I. Bars, R. Bijker, and F. Iachello,
Phys. Rev. C {\bf 27}, 1761 (1983).
\bibitem{bij88} R. Bijker and V.K.B. Kota, Ann. Phys. (N.Y.)
{\bf 187}, 148 (1988).
\bibitem{wol91} S. Wolfram, {\it Mathematica} (Addison-Wesley,
Redwood City, 1991).
\bibitem{lam93} G.J. Lampard, A.E. Stuchbery, H.H. Bolotin, and S. Kuyucak,
to be published.
\bibitem{van84} P. Van Isacker, A. Frank, and H.Z. Sun, Ann. Phys. (N.Y.)
{\bf 157}, 183 (1984).
\bibitem{bij85} R. Bijker and F. Iachello, Ann. Phys. (N.Y.)
{\bf 161}, 360 (1985).
\bibitem{war82} D.D. Warner {\it et al.}, Phys. Rev. C {\bf 26}, 1921 (1982).
\bibitem{sun83} H.Z. Sun {\it et al.}, Phys. Rev. C {\bf 27}, 2430 (1983);
{\bf 29}, 352 (1984); {\bf 31}, 1899 (1985).
\bibitem{ver83} M. Vergnes {\it et al.}, Phys. Rev. C {\bf 28}, 360 (1983);
{\bf 30}, 517 (1984); {\bf 36}, 1218 (1987).
\bibitem{bru85} A.M. Bruce {\it et al.}, Phys. Lett. {\bf B165}, 43 (1985).
\bibitem{mau86} A. Mauthofer {\it et al.}, Phys. Rev. C {\bf 34}, 1958 (1986);
{\bf 39}, 1111 (1989).
\bibitem{stu91} A.E. Stuchbery, G.J. Lampard, and H.H. Bolotin, Nucl. Phys.
{\bf A528}, 447 (1991).
\bibitem{rag89} P. Raghavan, At. Data Nucl. Data Tab.
{\bf 42}, 189 (1989).
\end{references}
\end{document}